\documentclass[aps,prl,twocolumn]{revtex4-1}

\usepackage{upgreek}
\usepackage{amsmath}
\usepackage{amsfonts}
\usepackage{amssymb}
\usepackage{graphicx}
\usepackage[colorlinks,linkcolor=blue,anchorcolor=blue,citecolor=blue,urlcolor=blue]{hyperref}
\usepackage{mathrsfs}
\usepackage{epsfig}
\usepackage{float}
\usepackage{subfigure}
\usepackage{mathtools}
\usepackage{url}

\begin{document}
\bibliographystyle{apsrev4-1}

\title{Vibrational Kerr solitons in an optomechanical microresonator}

\author{Jia-Chen Shi$^{1}$}
\author{Qing-Xin Ji$^{1}$}
 \altaffiliation[Current address: ]{T. J. Watson Laboratory of Applied Physics, California Institute of Technology, Pasadena, CA, USA.} 
\author{Qi-Tao Cao$^{1}$}
\author{Yan Yu$^{1}$}
 \altaffiliation[Current address: ]{T. J. Watson Laboratory of Applied Physics, California Institute of Technology, Pasadena, CA, USA.} 
\author{Wenjing Liu$^{1}$}
\author{Qihuang Gong$^{1,2}$}
\author{Yun-Feng Xiao$^{1,2}$}
\email{yfxiao@pku.edu.cn}

\affiliation{$^{1}$State Key Laboratory for Mesoscopic Physics and Frontiers Science Center for Nano-optoelectronics, School of Physics, Peking University, 100871, Beijing, China\\
$^2$Collaborative Innovation Center of Extreme Optics, Shanxi University, Taiyuan 030006, China}

\date{\today}

\begin{abstract}
Soliton microcombs based on Kerr nonlinearity in microresonators have been a prominent miniaturized coherent light source.
Here, for the first time, we demonstrate the existence of Kerr solitons in an optomechanical microresonator, for which a nonlinear model is built by incorporating a single mechanical mode and multiple optical modes.
Interestingly, an exotic vibrational Kerr soliton state is found, which is modulated by a self-sustained mechanical oscillation.
Besides, the soliton provides extra mechanical gain through the optical spring effect, and results in phonon lasing with a red-detuned pump.
Various nonlinear dynamics is also observed, including limit cycle, higher periodicity, and transient chaos.
This work provides a guidance for not only exploring many-body nonlinear interactions, but also promoting precision measurements by featuring superiority of both frequency combs and optomechanics.
\end{abstract}

\maketitle

Chip-integrable optical frequency combs in Kerr microresonators have attracted much interest in recent years \cite{del2007optical,kippenberg2011microresonator}. 
In particular, dissipative Kerr soliton (DKS) microcombs, resulted from the balance between dispersion and optical Kerr nonlinearity, provides a miniaturized coherent link between the optical and microwave frequencies \cite{herr2014temporal,kippenberg2018dissipative,diddams2020optical}. So far, DKSs have been demonstrated with prominent capabilities in a wide range of applications, such as optical clocks \cite{newman2019architecture,drake2019terahertz}, LiDARs \cite{riemensberger2020massively,suh2018soliton},
optical computations \cite{xu202111,feldmann2021parallel}, and high-resolution spectroscopy \cite{suh2016microresonator,yang2019vernier}.
Besides, DKSs have also been employed as an entity for investigating nonlinear physics, accompanied with various novel phenomena including breathing soliton \cite{bao2016observation,yu2017breather,lucas2017breathing}, soliton crystals \cite{cole2017soliton,lu2021synthesized}, and soliton bursts \cite{zhou2019soliton,yu2021spontaneous}. 

\begin{figure}[b]
\centering
\includegraphics[width=8.5cm]{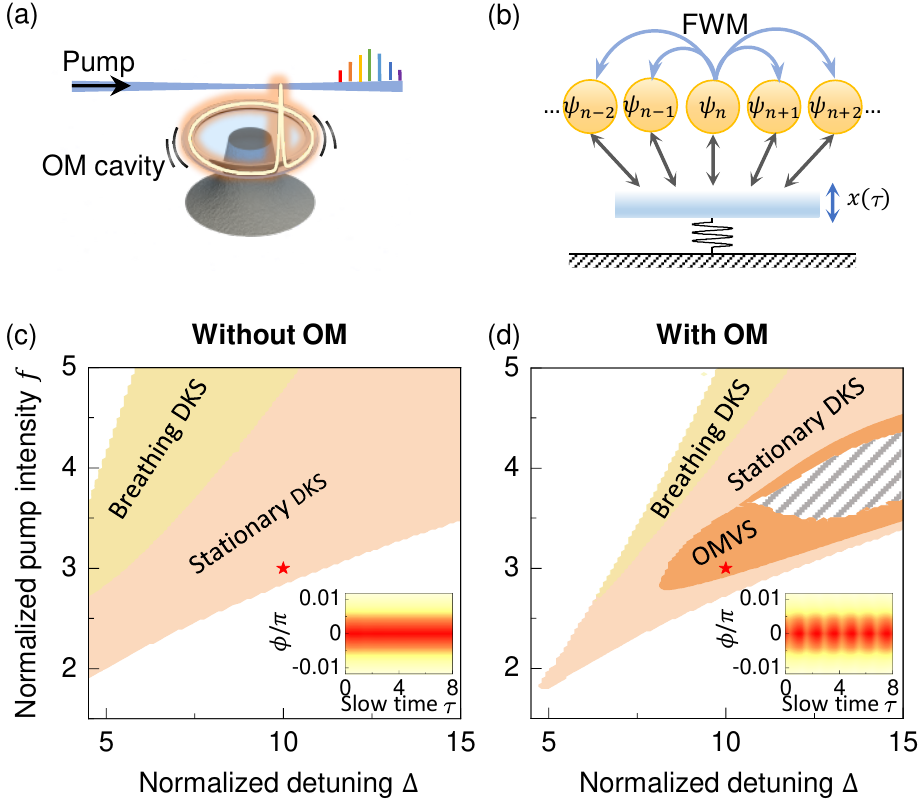}
\caption{
(a) Illustration of a DKS in a WGM optomechanical microresonator.
(b) Schematic diagram of coupled optical modes $\psi_n$ interacting with a single dissipative mechanical mode with displacement \textbf{$x(\tau)$}. FWM: four-wave-mixing. 
(c)-(d) Numerically simulated phase diagram of the coupled system without and with optomechanical (OM) interaction. Soliton exists in the colored region, with breathing DKS, stationary DKS, and OMVS.
Inset: Dynamical evolution of stationary DKS and OMVS over a normalized slow time (at $\Delta=10$, $f=3$ marked as the star) versus azimuthal angle $\phi$ over a single round trip.
Parameters:
$\beta=0.01$, $\omega_{\rm m}=5.17$, $\gamma_{\rm m}=0.0517$, $G=6.18\times 10^{11}$ m$^{-1}$, 
$C= 1.51 \times 10^{-10}$ m.
}
\label{fig:1}
\end{figure}

Very recently, it is found that the characteristics of temporal DKSs can be engineered by other nonlinear effects, such as stimulated scattering \cite{yi2015soliton,yang2016stokes,wang2018stimulated}, thermo-optic effect \cite{yang2021dispersive,leshem2021thermal,guo2017universal,stone2018thermal} and harmonic generations \cite{bruch2021pockels}. 

These effects provide strategies for manipulating microcomb properties, e.g., frequency extension through second harmonic generation and noise reduction through Brillouin scattering
\cite{guo2017universal,stone2018thermal,bruch2021pockels,bai2021brillouin}. 
Among various cavity-enhanced nonlinear effects, optomechanics is ubiquitous in ultrahigh-$Q$ microcavities and holds
unprecedented abilities in controlling macroscopic quantum states \cite{aspelmeyer2014cavity,schliesser2008resolved,verhagen2012quantum, ma2021non,magrini2021real} as well as
precision measurements \cite{bagci2014optical,arcizet2006radiation, thompson2008strong, brawley2016nonlinear,aggarwal2020room}.
Although Kerr microcombs are observed preliminarily in optomechanical microresonators \cite{chen2020chaos,zhang2020spectral,suzuki2017suppression}, the existence of DKSs under the strong optomechanical coupling is still elusive.

In this Letter, we demonstrate the existence of robust DKSs in a strongly coupled optomechanical microresonator. By incorporating the interaction between a single mechanical mode and multiple optical modes into the nonlinear Schrödinger equation,
a vibrational state of solitons is found analytically, which is modulated by the mechanical oscillation.
Moreover, it is revealed that the soliton provides extra gain to the mechanical resonator through the optical spring effect and leads to phonon lasing working with a red-detuned optical pump.
Rich nonlinear dynamics are also observed for the optomechanically vibrational solitons, including limit cycle, periodic doubling and transient chaos. 
This work paves the way for exploring complex nonlinear dynamics, as well as potential applications of precision measurements by combining frequency combs and optomechanics.

As shown in Fig. \ref{eqn:1}(a), a whispering gallery mode (WGM) optomechanical microresonator is adapted for DKS generation, where the cavity boundary vibrates due to the enhanced light-radiation pressure \cite{aspelmeyer2014cavity}. 
Under a red-detuned pump input, the cavity modes $\psi_\mathrm{n}$ 
forming the soliton combs couple to each other through Kerr four-wave-mixing (FWM), and interact with a single mechanical mode, as illustrated in Fig. \ref{fig:1}(b). 
The evolution of this coupled system is described by the Lugiato-Lefever equation \cite{lugiato1987spatial} incorporated with optomechanical coupling \cite{aspelmeyer2014cavity},
\begin{equation}
    \frac{\partial\psi(\tau,\phi)}{\partial \tau}    =i\frac{\beta}{2}\frac{\partial^2\psi}{\partial\phi^2}+i(G x+\vert\psi\vert^2-\Delta)\psi-\psi+f,
    \label{eqn:1}
\end{equation}
\begin{equation}
    \frac{\mathrm{d}^2x}{\mathrm{d}\tau^2}+\gamma_{\rm m}\frac{\mathrm{d}x}{\mathrm{d}\tau}+\omega_{\rm m}^2x=C\sum_{\rm n}\vert\psi_{\rm n}\vert^2,
    \label{eqn:2}
\end{equation}
where $\psi(\tau,\phi)=\sum_{\rm n} \psi_{\rm n} e^{i n \phi} $ is the normalized intracavity waveform dependent on the azimuthal angle $\phi$ and normalized evolution time $\tau$.
$x(\tau)$ denotes the displacement of the mechanical mode with resonance frequency $\omega_{\rm m}$, decay rate $\gamma_{\rm m}$ and radiation pressure coefficient $C$. 
The terms on the right-hand side of Eq. \ref{eqn:1} describe, respectively, group-velocity dispersion with the dispersion parameter $\beta$, mechanical back action with coefficient $G$, Kerr nonlinearity, frequency detuning between the central optical mode and the pump laser $\Delta=\omega_0-\omega_{\rm p}$, cavity losses, and external driving $f$ on a single optical mode $\psi_0$.
The mechanical resonant frequency $\omega_\mathrm{m}$ is larger than the optical decay rate (the resolved-sideband regime),
and it is assumed that the optical modes share the same coupling strength $G$ with the mechanical mode.
All these coefficients are normalized to the cavity photon lifetime \cite{supple}. 
\begin{figure}[t]
\centering
\includegraphics[width=8.5cm]{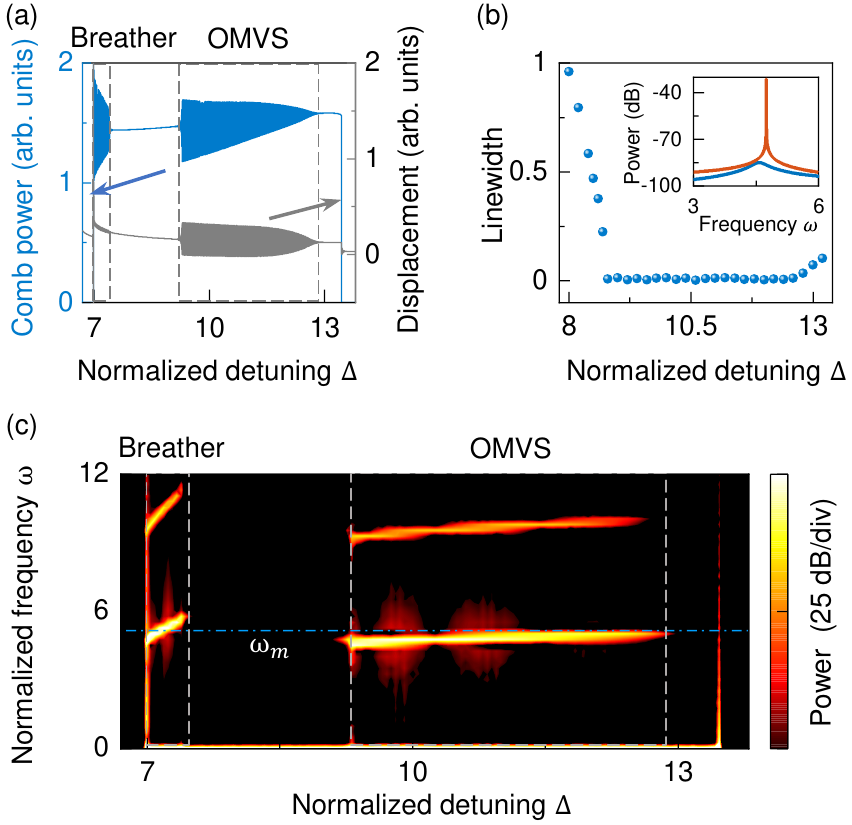}
\caption{
(a) Numerical simulated trace of comb power (blue) and the displacement of the mechancial resonator normalized by 10 pm (grey) over the laser detuning scan starting from a stationary DKS state at $f=3.2$. 
(b) Numerical linewidth of the mechanical oscillator under different detuning. Inset: frequency spectra of mechanical resonator at $\Delta=8.4$ (blue) and $\Delta=10$ (orange).
(c) The frequency spectra over the laser detuning. The blue dashed-dotted line indicates the mechanical frequency.
The grey dashed lines in a,c indicate different regions.
}
\label{fig:2}
\end{figure}

\begin{figure*}[t]
\centering
\includegraphics[width=14cm]{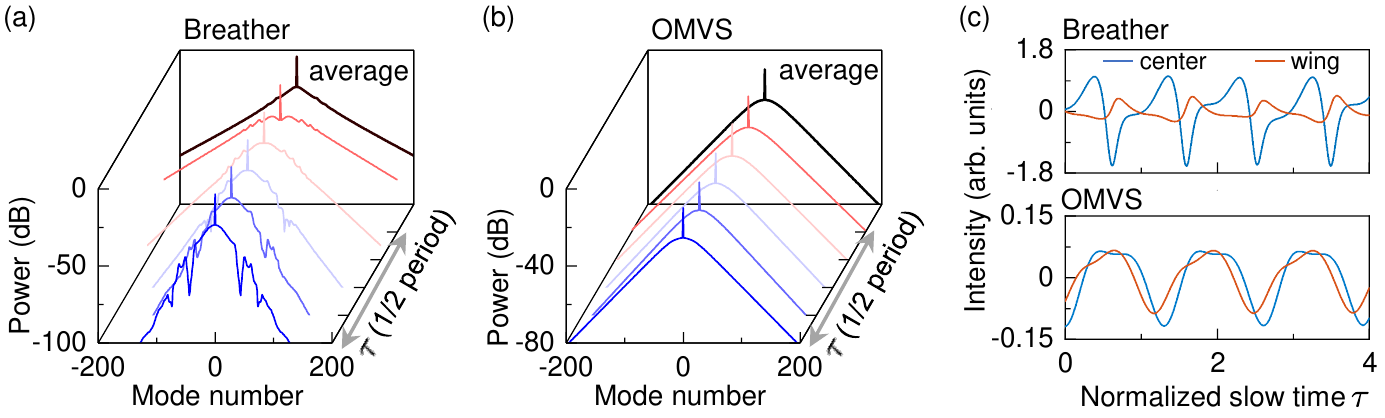}
\caption{
(a)-(b) Evolution of optical spectra  for the breathing solitons at $\Delta=10$, $f=5$ and OMVS at $\Delta=10$, $f=3$, respectively. The red curves are the average spectrum within a half period. 
(c) The corresponding evolution of power amplitude of the central comb lines (blue) and the lines in the wing (orange) for the breathing solitons.
}
\label{fig:25}
\end{figure*}

In absence of the optomechanical effects, the existence of the soliton states depends on both the continuous wave (CW) pump intensity $f$ and the frequency detuning $\Delta$, and these solitons can be classified into stationary solitons \cite{herr2014temporal} and breathing solitons, i.e., breathers, which exhibits periodic oscillations \cite{bao2016observation,yu2017breather,lucas2017breathing}, as shown in Fig. \ref{fig:1}(c). 
Once strong optomechanical coupling is introduced, 
by solving Eq. (\ref{eqn:1}), (\ref{eqn:2}) with typical experimental parameters \cite{schliesser2008resolved,suzuki2017suppression},
it is found that single DKS still exists within certain regions, of which all the boundaries redshift compared with the case without optomechanical effects, as shown in Fig. \ref{fig:1}(d).
This is because the intracavity optical field increases the equilibrium cavity length of the mechanical resonator, results in an overall red-shift of the cavity modes in turn \cite{aspelmeyer2014cavity}.
It is also noticed that the soliton vanishes within the shaded area in Fig. \ref{fig:1}(d), which will be explained later.

In the optomechanical microcavity, a new type of soliton state emerges, characterized by a temporal oscillation in its intensity, which is referred to as the optomechaincally vibrational soliton (OMVS), as shown in the inset of Fig. \ref{fig:1}(d).
As both breathers and OMVSs possess similar oscillatory nature, their differences seem subtle, however, they originate from fundamentally different mechanisms.
To illustrate their physical origin, the normalized detuning is adiabatically scanned at a constant pump intensity $f=3.2$, thus that the system is consecutively switched between different soliton states, as shown in Fig. \ref{fig:2}(a).
Although oscillating comb powers are observed in both the breathing soliton and OMVS regimes, the mechanical resonator vibration is much stronger in the OMVS than in the breather.
Spectrum-wise, during the laser scanning, the oscillation frequency of OMVS asymptotically approaches the mechanical frequency $\omega_{\rm m}$ in the entire oscillating regime due to optical spring effect (Fig. \ref{fig:2}(c)) \cite{aspelmeyer2014cavity}.
On the contrary, the breather oscillating frequency grows linearly with detuning independent on the mechanical frequency, consistent with previously reported breathers in absence of the optomechanical interaction \cite{lucas2017breathing}. 
These results imply that the OMVS originates from the strong interaction between the optical modes and mechanical modes, whereas the breather is less affected by the optomechanical coupling. 
Interestingly, approaching the onset of the OMVS, the linewidth of the mechanical resonance narrows down rapidly and nearly vanishes at the threshold, as shown in Fig. \ref{fig:2}(b).
It indicates the emergence of self-sustained mechanical oscillation in the OMVS regime, also known as phonon lasing.

For a direct illustration of the soliton spectral dynamics, the evolution of the mode profiles of the breather and OMVS states with typical parameters (breather: $\Delta=10$, $f=3$ and OMVS: $\Delta=10$, $f=5$) are compared in Fig. \ref{fig:25}.
As for the breather,
the energy oscillates between the center and wing comb lines, giving rise to a oscillating optical spectral envelop that deviates from the stationary soliton state (Fig. \ref{fig:25}(a)). 
The time averaged spectrum is characterized by a triangular shape in the logarithmic scale, analogous to the conventional Kuznetsov–Ma breathing solitons \cite{bao2016observation,yu2017breather,lucas2017breathing}.
Oppositely, the whole OMVS envelop oscillates rigidly as a particle-like wave package while maintaining the $\mathrm{sech^2}$ shaped envelop, which is typical for stationary solitons.
Another evidence for revealing their different oscillation origins is by examining the relative phases between the center and wing comb lines.
in the breather, the center comb lines (mode number $n=1 \sim 5$) oscillate in nearly opposite phases with those in the wings (mode number $n=21 \sim 25$), thus implying intra-comb energy exchange \cite{bao2016observation,lucas2017breathing} (Fig. \ref{fig:25}(c), upper). 
Contrarily, all the comb lines in the OMVS spectrum oscillate in almost synchronized phases, indicating their coherent energy transfers with the mechanical mode (Fig. \ref{fig:25}(c), lower).
Such different energy highlights the potential advantages of OMVSs in e.g., sensing applications, as they are strongly coupled to a mechanical oscillator while preserving the intrinsic stability of solitons.

\begin{figure}[b]
\centering
\includegraphics[width=8.5cm]{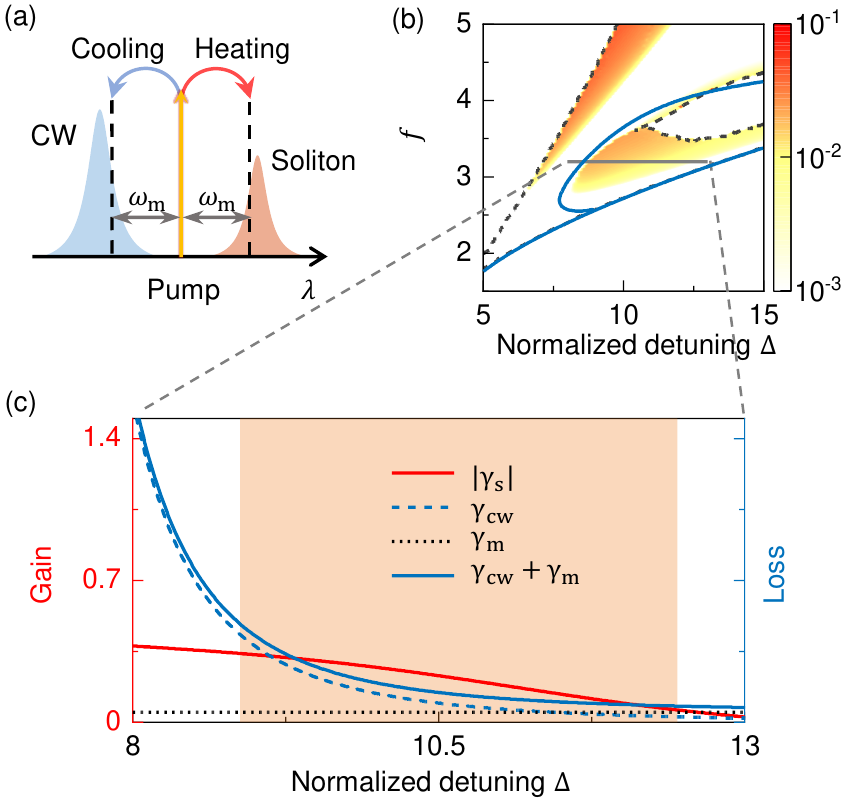}
\caption{
(a) Scheme of cooling and heating of the mechanical oscillator with optical radiation pressure. 
(b) Numerical oscillation amplitude for comb power (color) (upper panel). The area encircled by blue curve is the vibrational soliton region based on the Lagrangian model.
(c) Gain and loss of mechanical mode at the pump intensity $f=3.2$ corresponding to the gray line in (b). The shaded denotes the numerical OMVS region with manifested oscillation (lower panel).
}
\label{fig:4}
\end{figure}

Notably, in conventional optomechanical systems, the self-sustained mechanical oscillation oscillator is well known to exists only with the blue-sideband drive, whereas in our system, it is enabled with the drive at the red sideband \cite{marquardt2006dynamical, bakemeier2015route,carmon2005temporal,carmon2007chaotic}. 
In absence of the DKS formation, the photons from the red-detuned pump laser can be scattered by absorbing phonons through the cavity-enhanced anti-Stokes process, resulting in the cooling of the mechanical resonator (i.e., suppressed oscillator vibration) \cite{kippenberg2005analysis,schliesser2008resolved}.
Interestingly, once the DKS is generated, the intracavity soliton pulse can shift part of the mode resonance to the red side of the input laser, forming an additional resonance by the Kerr effect, {\it i.e.}, the $\mathcal{S}$-resonance  \cite{guo2017universal}.
The pump laser is effectively blue-detuned to this new resonance, which thus introduces extra gain to the mechanical oscillator by releasing phonons through the enhanced Stokes process, as schematically illustrated in Fig. \ref{fig:4}(a).
In this case, the total loss of the mechanical resonator, including both the intrinsic and the optomechanical loss induced by the anti-Stokes process, can be compensated by the gain from the $\mathcal{S}$-resonance, enabling the self-sustained mechanical oscillation.

Moreover, the Lagrangian method is employed for analytically studying the emergence of OMVS, as shown in Fig. \ref{fig:4}(b).
Through identifying the sign of the real part of the leading eigenvalue, the threshold for the self-sustained oscillation is obtained, which is plotted in blue in Fig. \ref{fig:4}(b) and agrees well with the numerical boundary of teh OMVS region \cite{supple}.
Furthermore, the different contributions of mechanical loss/gain is extracted under large detuning approximation $\Delta^2 \gg 1$, and the overall loss of the mechanical resonator reads,
\begin{equation}
    \gamma=\gamma_{\rm m}+\gamma_{\rm cw}+\gamma_{\rm s}, 
\end{equation}
where $\gamma_{\rm m}$ is the intrinsic loss and $\gamma_{\rm cw}$ ($\gamma_{\rm s}$) is the contribution through anti-Stokes (Stokes) scattering from the CW background (soliton).
Here the optomechanical damping rate is derived from the imaginary part of the mechanical susceptibility which is obtained by calculating the mechanical response to the radiation-pressure force \cite{supple,kippenberg2005analysis,aspelmeyer2014cavity}.
Note that when the DKS is generated, pump laser is effectively blue detuned to the $\mathcal{S}$-resonance corresponding to negative term $\gamma_{\rm s}$, and the loss of the mechanical oscillator is hereby reduced.
For the pumping conditions marked by the gray line in Fig. \ref{fig:4}(c), upon the increase of the detuning $\Delta$, both the optomechanical loss $\gamma_{\rm cw}$ and gain $\vert\gamma_{\rm s}\vert$ drop with different variation rates.
When the mechanical gain exactly compensates the total loss, the threshold of phonon lasing is reached, which agrees well with the numerical boundary of the OMVS illustrated by the orange shaded region.
Once the gain exceeds the loss, the net loss becomes negative, leading to an exponentially amplified oscillation until saturation \cite{aspelmeyer2014cavity}.

\begin{figure}[!t]
\centering
\includegraphics[width=8cm]{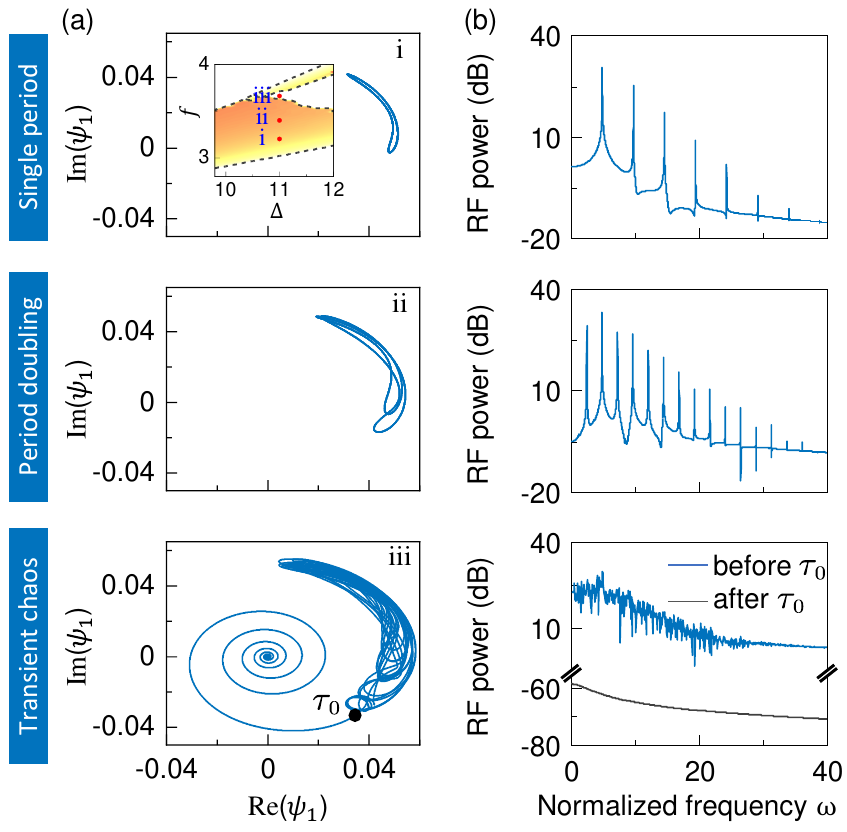}
\caption{Periodicity and chaotic evolution of the OMVS.
(a) Evolution of the real and imaginary part of the electricity field of the second comb line with $\Delta=11$.
Point i: single period at $f=3.2$; point ii: period doubling at $f=3.4$; ponit iii: transient chaos at $f=3.64$. Inset: their corresponding coordinate in the phase diagram.
(b) Corresponding RF spectra. 
The blue (gray) curve indicates the spectrum before (after) $\tau_0$.
}
\label{fig:5}
\end{figure}

Finally, rich temporal dynamics can be revealed in the evolution of OMVS states due to the strong nonlinearity.
As shown in point i of Fig. \ref{fig:5}, a distorted limit-cycle structure is observed with the pumping conditions marked by the red dots in the inset in the phase space spanned by the real and imaginary parts of the optical field of the comb line $\psi_{\rm 1}$.
The corresponding frequency spectrum exhibits multiple harmonics, attributed to cascaded phonon scattering and nonlinearity of the mechanical oscillation.
As the pump intensity increases at a constant detuning, a period doubling bifurcation emerges, with a limit cycle in the phase space with period twice as in i (point ii). 
This period doubling is also confirmed by the frequency spectrum, where new peaks present at half of the initial resonant frequencies. 
Under a higher driving power (point iii), the optical field turns into an irregular evolution before a moment $\tau_{\rm 0}$ with chaotic orbits in the phase space, exhibiting a continuous and highly noisy frequency spectrum. 
This transient chaos is induced by the optomechanical nonlinearity through switching energy between the multi-sidebands, and the original state becomes unstable once the chaotic orbit enters the attractive region of an adjacent basin \cite{leo2013dynamics,bakemeier2015route}.
Hence, after the critical moment $\tau_{\rm 0}$, the soliton exits from the limit cycle of the OMVS and eventually collapses to the CW-background state.

In summary, we have demonstrated the existence of DKS in a microcavity with strong optomechanical interactions.
By building a nonlinear model incorporating a single mechanical mode and multiple optical modes, a new form of soliton appears with the periodical oscillations, which is modulated by a self-sustained mechanical oscillation.
Besides, the soliton provides extra mechanical gain through the optical spring effect  and results in phonon lasing with a red-detuned pump.
We also observe various nonlinear dynamical orbits of the OMVSs, containing limit cycle, higher periodicity and transient chaos.
This work lays the groundwork for operating soliton microcombs in microcavities with strong optomechanical coupling and provides applications, e.g. frequency division and random bits generation \cite{cole2019subharmonic,okawachi2016quantum}. 
Moreover, the incorporation of frequency combs and optomechanics, two powerful tools for ultra-sensitive measurements, holds great potential for developments in precision metrology and trace detection \cite{newman2019architecture,drake2019terahertz,riemensberger2020massively,suh2018soliton,bagci2014optical,arcizet2006radiation, thompson2008strong,brawley2016nonlinear,aggarwal2020room}.

We thank Q.-F. Yang, Y.-C. Liu and H.-J. Chen at Peking University for helpful discussions. This project is supported by the National Key R$\&$D Program of China (Grants 2018YFB2200401), and the National Natural Science Foundation of China (Grants No. 12041602, 11825402, 11654003, and 62035017). Q.-T. Cao is supported by the National Postdoctoral Program for Innovative Talents (Grant BX20200014) and the China Postdoctoral Science Foundation (Grant 2020M680185).


\bibliography{apssamp}

\end{document}